\documentclass[manuscript,screen]{acmart}

\newlength\mylen
\usepackage{algorithm}
\usepackage{algpseudocode}
\usepackage{tabularx}
\usepackage{tikz}
\usepackage{array}
\newcolumntype{C}{>{\hfil$}p{\mylen}<{$\hfil}}
\usetikzlibrary{matrix}

\newcommand{\cdiv}[1]{\ensuremath{\mbox{cdiv}(#1)}}

\newcommand{\glbind}[1]{\ensuremath{\mbox{glbind}(#1)}}

\newcommand{\indmap}[1]{\ensuremath{\mbox{indmap}(#1)}}
\newcommand{\indmapn}{$\mbox{indmap}(1$:$n)$}

\newcommand{\Lnz}[1]{\ensuremath{\mbox{Lnz}(#1)}}
\newcommand{\nul}{\ensuremath{\mbox{\bf null}}}

\newcommand{\relind}[1]{\ensuremath{\mbox{relind}(#1)}}
\newcommand{\relindB}[1]{\ensuremath{\mbox{relindB}(#1)}}

\newcommand{\tn}{$t(1$:$n)$}

\makeatletter
\algnewcommand{\LineComment}[1]{\Statex \hskip\ALG@thistlm /*\ #1\ */}
\newcommand{\multiline}[1]{%
	\begin{tabularx}{\dimexpr\linewidth-\ALG@thistlm}[t]{@{}X@{}}
		#1
	\end{tabularx}
}
\makeatother
\algnewcommand{\IIf}[1]{\State\algorithmicif\ #1\ \algorithmicthen}
\algnewcommand{\EndIIf}{\unskip\ \algorithmicend\ \algorithmicif}
\algnewcommand{\FFor}[1]{\State\algorithmicfor\ #1\ \algorithmicdo}
\algnewcommand{\EndFFor}{\unskip\ \algorithmicend\ \algorithmicfor}
\algrenewcommand\algorithmicindent{1.0em}%
\algnewcommand{\IfThenElse}[3]{
\State \algorithmicif\ #1\ \algorithmicthen\ #2\ \algorithmicelse\ #3}

\begin{document}

\title{Some new techniques to use in serial sparse {C}holesky factorization algorithms}

\author{M.  Ozan Karsavuran}
\affiliation{%
  \institution{Lawrence Berkeley National Laboratory}
  \country{USA}
}
\author{Esmond G. Ng}
\affiliation{%
  \institution{Lawrence Berkeley National Laboratory}
  \country{USA}
}

\author{Barry W. Peyton}
\affiliation{%
  \institution{Dalton State College}
  \country{USA}
}

\author{Jonathan L. Peyton}
\affiliation{%
  \institution{Intel Corporation}
  \country{USA}
}


\begin{abstract}
We present a new variant of serial right-looking supernodal sparse 
Cholesky factorization (RL).
Our comparison of RL with the multifrontal method confirms
that RL is simpler, slightly faster, 
and requires slightly less storage.
The key to the rest of the work in this paper is recent work on reordering columns
within supernodes so that the dense off-diagonal blocks in the factor
matrix joining pairs of supernodes are fewer and larger.
We present a second new variant of serial right-looking
supernodal sparse Cholesky factorization (RLB), 
where this one is specifically designed to exploit
fewer and larger off-diagonal blocks in the factor matrix
obtained by reordering within supernodes.
A key distinction found in RLB is that it
uses no floating-point working storage
and performs no assembly operations.
Our key finding is that RLB is unequivocally faster than its competitors.
Indeed, RLB is consistently, but modestly, faster than its competitors
whenever Intel's MKL {\em sequential} BLAS are used. 
More importantly,
RLB is substantially faster than its competitors whenever 
Intel's MKL {\em multithreaded} BLAS are used.
Finally, RLB using the multithreaded BLAS achieves impressive speedups over
RLB using the sequential BLAS.
\end{abstract}





\maketitle

%
%

\section{Introduction}
\label{sec:intro}

Writing well-designed and reliable software that solves large sparse
symmetric positive definite linear systems via sparse Cholesky factorization
has always been challenging enough that it has been
carried out by specialists in the field of sparse-matrix computations.
Indeed, the motivation for this work comes more from software-related considerations
than from algorithm-related considerations,
though new algorithms are involved.
We are convinced that it is well worthwhile
to explore and develop serial sparse Cholesky factorization
algorithms to the fullest extent possible in light of several developments
that have occurred since such algorithms
were last a focus of any serious research.
With the desire to preserve as much simplicity as possible in the
algorithms and the software,
our aim is to see how well serial sparse Cholesky algorithms
can exploit parallelism (i.e., multiple cores) via the multithreaded BLAS
routines that they invoke.
In essence, we are exploring how far one can get by applying
to sparse Cholesky factorization
the same techniques and methodology used to parallelize LAPACK.



Let~$A$ be an $n$~by~$n$ sparse symmetric positive definite matrix, and
let $A=LL^T$ be the Cholesky factorization of~$A$,
where~$L$ is a lower triangular matrix.
It is well known that~$L$ suffers {\em fill} during such a factorization;
that is, $L$ will have nonzero entries in locations occupied
by zeros in~$A$.
As a practical matter, it is important to limit the number of such fill entries
in~$L$.
It is well known that solving a sparse symmetric positive definite
linear system $Ax=b$ via sparse Cholesky factorization requires 
four steps~\cite{GL81}:
\begin{enumerate}
  \item
    {\bf (Ordering)} Compute a fill-reducing ordering of~$A$ using either the 
    {\em nested dissection}~\cite{George73,KK99}
    or the
    {\em minimum degree}~\cite{ADD96,GL89,Liu-MMD-1985,Tinney-Walker67}
    ordering heuristic.
  \item
    {\bf (Symbolic factorization)} Compute the needed information about and data structures for
    the sparse Cholesky factor matrix.
  \item
    {\bf (Numerical factorization)} Compute the sparse Cholesky factor within the data structures
    computed in step~2.
  \item
    {\bf (Solution)}  Solve the linear system by performing in succession
    a sparse forward solve and a sparse backward solve using
    the sparse Cholesky factor and its transpose, respectively.
\end{enumerate}
Our primary focus in this paper is on serial algorithms for
numerical factorization.
It will be convenient henceforth to assume that the matrix~$A$
has been reordered by the fill-reducing ordering obtained in step~1,
so that $A=LL^T$.

The key to the first truly efficient algorithms for sparse
Cholesky factorization is the existence of {\em supernodes}
in the factor matrix.
Loosely speaking, a {\em supernode} is a set of consecutive columns
in the factor matrix that share the same zero-nonzero structure.
Because of this shared structure, each supernode can be treated
as a dense submatrix in certain efficient sparse Cholesky factorization
algorithms.
The first of these algorithms was the multifrontal method~(MF) 
introduced in 1983 by Duff and Reid~\cite{DR83}.
The multifrontal method is probably currently the most commonly 
used factorization method,
so it is natural to include it in our study.

A second method that exploits supernodes in a similar way
was introduced independently by
Rothberg and Gupta~\cite{RG91} and Ng and Peyton~\cite{NP94}.
In this method, the computation is organized so that the updates
to the current supernode come from supernodes to the left.
Hence, we will refer to it as the {\em left-looking} method~(LL).
This method will also be included in our study.

The remaining two factorization methods are new, though they
can be viewed as variants derived from predecessors.
The first of these can be viewed as a simplification of MF.
In the multifrontal method, when a supernode is computed,
what ultimately happens to its updates for supernodes to the right
is fairly complicated, as we shall see.
We have created a {\em right-looking} factorization method~(RL),
where the updates are incorporated in a simple fashion
into the appropriate supernodes to the right.
In our study, we will document that~RL is simpler than~MF,
and that it modestly reduces the time and storage required relative
to~MF.

There is one feature common to all three of the factorization
methods~MF, LL, and~RL.
Each uses a piece of floating-point working storage to accumulate
updates from a completed supernode,
and subsequently {\em assembles} 
(i.e., scatter-adds) them into the target to receive the updates.
As we shall see, this shared feature of assembling updates, 
which exists specifically to handle sparsity issues,
creates some performance difficulties for each of these three methods.
Our fourth (and final) factorization method performs no such
assembly operations;
this fourth method is the primary contribution of this paper.

The fourth factorization method can be viewed as a modification
of~RL, and we will refer to the method as {\em right-looking blocked}
(RLB).
We will next introduce RLB by sketching out its application
to the small sparse Cholesky factor shown in Figure~\ref{fig:supernode1}.
Before doing so, we will need the following notation as we examine the figure;
for a matrix~$C$ and two index sets~$K$ and~$J$ we will let
$C_{K,J}$ be the submatrix of~$C$ with rows taken from~$K$
and columns taken from~$J$.

\begin{figure}[htp]
\footnotesize
\setlength{\arraycolsep}{3pt}
\begin{center}
      \begin{eqnarray*}
      {\color{white} L} & {\color{white} =} &
      {\color{white} \left[
       \begin{array}{cc|cc|ccccc}
         \multicolumn{2}{c}{\color{black} J_1} &
         \multicolumn{2}{c}{\color{black} J_2} &
         \multicolumn{5}{c}{\color{black} J_3} \\
           {\color{white} 1} & 
           {\color{white} +} & 
           {\color{white} 3} & 
           {\color{white} +} & 
           {\color{white} +} & 
           {\color{white} +} & 
           {\color{white} +} & 
           {\color{white} 8} & 
           {\color{white} 9}
       \end{array}
       \right]} \\
       L & = &
       \left[
       \begin{array}{cc|cc|ccccc}
         {1} & & \multicolumn{7}{c}{} \\
         \ast & {2} & \multicolumn{7}{c}{} \\
         & & 3 & & \multicolumn{5}{c}{} \\
         & & \ast & 4 & \multicolumn{5}{c}{} \\
         \ast & \ast & \ast & \ast & 5 & \multicolumn{4}{c}{} \\
         \ast & + & & & \ast & 6 & \multicolumn{3}{c}{} \\
         & & \ast & + & + & \ast & 7 & \multicolumn{2}{c}{} \\
         & & \ast & \ast & \ast & + & \ast & 8 & \multicolumn{1}{c}{} \\
         \ast & \ast & & & + & \ast & + & \ast & 9
       \end{array}
       \right]
      \end{eqnarray*}
  \caption{The supernodes of a sparse Cholesky factor $L$.  
	Each symbol~`$\ast$' signifies
	an off-diagonal entry that is nonzero in both~$A$ and~$L$; 
	each symbol~`$+$' signifies an off-diagonal entry that is zero in~$A$ but nonzero
	in~$L$---a fill entry in~$L$.}
  \label{fig:supernode1}
\end{center}
\end{figure}

In the figure, the first supernode $J_{1} = \{1,2\}$ comprises columns $1$ and $2$.
The $2$~by~$2$ submatrix $L_{J_{1},J_{1}}$ is a dense lower triangular matrix.
Below this dense lower triangular block, columns $1$ and $2$ of $L$
share the same sparsity structure;
specifically, both have nonzeros in rows $5$,~$6$, and~$9$ only.
The second supernode $J_{2} = \{3,4\}$ comprises columns $3$ and $4$.
The $2$~by~$2$ submatrix $L_{J_{2},J_{2}}$ is a dense lower triangular matrix.
Below this dense lower triangular block, columns $3$ and $4$ of $L$
share the same sparsity structure;
specifically, both have nonzeros in rows $5$,~$7$, and~$8$ only.
The third supernode $J_{3} = \{5,6,7,8,9\}$ comprises columns $5$ through $9$.
The $5$~by~$5$ submatrix $L_{J_{3},J_{3}}$ is a dense lower triangular matrix.

To compute~$L$, the algorithm RLB first computes supernode~$J_{1}$, 
which requires no updates from the left.
Supernode~$J_{1}$ has two dense {\em blocks} joining it to 
supernode~$J_{3}$, namely, 
$B=\{5,6\}$ and $B'=\{9\}$,
where each set identifies the rows in the block.
We exploit these blocks in
RLB, which now updates supernode~$J_{3}$ with supernode~$J_{1}$,
as follows.
It first computes the lower triangle of the following:
\[
  L_{B,B} \gets L_{B,B} - L_{B,J_{1}} * L_{B,J_{1}}^T
\]
using the BLAS routine DSYRK.
It then computes
\[
  L_{B',B} \gets L_{B',B} - L_{B',J_{1}} * L_{B,J_{1}}^T
\]
using the BLAS routine DGEMM.
It then computes the lower triangle of the following:
\[
  L_{B',B'} \gets L_{B',B'} - L_{B',J_{1}} * L_{B',J_{1}}^T
\]
using the BLAS routine DSYRK.
Since $L_{J_2,J_1}=0$,
there are no updates from supernode~$J_{1}$ to supernode~$J_{2}$.

After supernode~$J_{1}$ is processed,
RLB will complete supernode~$J_{2}$.
Supernode~$J_{2}$ has two dense {\em blocks} joining it to 
supernode~$J_{3}$, namely, 
$B=\{5\}$ and $B'=\{7,8\}$.
At this point, supernode~$J_{2}$ will update supernode~$J_{3}$
with these blocks in the same way that supernode~$J_{1}$ did with
its blocks.
After that,
supernode~$J_{3}$ has received all of its updates,
and RLB finishes the computation by completing supernode~$J_{3}$.

Intuitively, the performance of~RLB would improve if the blocks could
somehow be made fewer and larger, but without increasing fill.
It turns out that we can achieve that goal by reordering the columns
within supernodes.

Let us turn our attention again to the sparsity pattern of the
Cholesky factor~$L$ shown in Figure~\ref{fig:supernode1}.
Recall that the supernodes were identified as
$J_{1} = \{1,2\}$,
$J_{2} = \{3,4\}$, and
$J_{3} = \{5,6,7,8,9\}$.
Let us now symmetrically permute the rows and columns of supernode $J_{3}$.
Specifically,
let us move row/column 6 to row/column 5 (i.e., $6 \rightarrow 5$),
along with $9 \rightarrow 6$,
$5 \rightarrow 7$,
$7 \rightarrow 8$, and
$8 \rightarrow 9$.
The sparsity pattern of the new Cholesky factor~$\widehat{L}$
is shown in Figure~\ref{fig:supernode2}.
\begin{figure}[htp]
\footnotesize
\setlength{\arraycolsep}{3pt}
\begin{center}
  \begin{eqnarray*}
    {\color{white} \widehat{L}} & {\color{white} =} &
    {\color{white} \left[
    \begin{array}{cc|cc|ccccc}
      \multicolumn{2}{c}{\color{black} J_1} &
      \multicolumn{2}{c}{\color{black} J_2} &
      \multicolumn{5}{c}{\color{black} J_3} \\
      {\color{white} 1} & 
      {\color{white} +} & 
      {\color{white} 3} & 
      {\color{white} +} & 
      {\color{white} +} & 
      {\color{white} 6} & 
      {\color{white} +} & 
      {\color{white} +} & 
      {\color{white} 9}
    \end{array}
    \right]} \\
    \widehat{L} & = &
    \left[
      \begin{array}{cc|cc|ccccc}
        $1$ & & \multicolumn{7}{c}{} \\
        \ast & $2$ & \multicolumn{7}{c}{} \\
        & & $3$ & & \multicolumn{5}{c}{} \\
        & & \ast & $4$ & \multicolumn{5}{c}{} \\ 
        \ast & + & & & {6} & \multicolumn{4}{c}{} \\
        \ast & \ast & & & \ast & {9} & \multicolumn{3}{c}{} \\ 
        \ast & \ast & \ast & \ast & \ast & + & {5} & \multicolumn{2}{c}{} \\
        & & \ast & + & \ast & + & + & {7} & \multicolumn{1}{c}{} \\
        & & \ast & \ast & + & \ast & \ast & \ast & {8} \\
      \end{array}
    \right]
  \end{eqnarray*}
  \caption{The supernodes of the sparse Cholesky factor $\widehat{L}$
  obtained after a symmetric permutation of supernode~$J_3$ in Figure~\ref{fig:supernode1}.
  Let $\widehat{A}$ be the new version of $A$ after the symmetric permutation.
  Each symbol~`$\ast$' signifies
  an off-diagonal entry that is nonzero in both~$\widehat{A}$ and~$\widehat{L}$; 
  each symbol~`$+$' signifies an off-diagonal entry that is zero in~$\widehat{A}$
  but nonzero in~$\widehat{L}$.}
  \label{fig:supernode2}
\end{center}
\mbox{}
\end{figure}

Consider now what happens when~RLB is used to compute~$\widehat{L}$.
Supernode~$J_{1}$ now has only one dense block joining it to
supernode~$J_{3}$.
As a consequence, there is now  only one block update from supernode~$J_{1}$
to supernode~$J_{3}$, rather than the three that were required under
the original ordering.
The reader should verify that the same is also true for block updates
from supernode~$J_{2}$ to supernode~$J_{3}$.

Recent research has led to excellent algorithms for reordering
columns within supernodes to reduce the number of blocks in this fashion.
The first truly successful algorithm of this sort
was introduced by Pichon, Faverge, Ramet, and Roman~\cite{PFRR17}.
They formulated the underlying combinatorial optimization problem
as a {\em traveling salesman problem};
hence, we will refer to their method as~TSP.
The problem with their approach was not ordering quality;
it was the cost, in time, of computing the needed~TSP distances~\cite{JNP18,PFRR17}.
Jacquelin, Ng, and Peyton~\cite{JNP21} devised a much faster way to 
compute the needed distances, which greatly reduces the runtimes
for the~TSP method.

Jacquelin, Ng, and Peyton~\cite{JNP18} proposed a simpler heuristic
for reordering within supernodes based on partition refinement~\cite{PT87}.
In their paper, they report faster runtimes for their method than~TSP,
while obtaining similar ordering quality.
We will refer to their method as~PR.

In a paper in progress~\cite{KNP24}, three of the authors
present a few incremental improvements to both~TSP and~PR.
Throughout this study, we will use~PR exclusively since it can be 
computed so quickly with no compromise in ordering quality;
this is based on comparisons of
our best and most recent implementations of the two methods,
as will be documented in~\cite{KNP24}.

Our testing indicates that~RLB, preceeded by PR, is an excellent 
serial sparse Cholesky factorization algorithm.
In our testing,
we use Intel's MKL sequential BLAS and Intel's MKL multithreaded BLAS;
we use the latter on~48~cores on our machine used for testing.
RLB is modestly superior to its competitors using the sequential BLAS,
but it is far better than its competitors using the multithreaded BLAS.

Our presentation will be expansive;
we wish to present all of the algorithms in detail,
and we wish, as much as possible,
for the paper to be self-contained.
Our paper is organized as follows.
In Section~\ref{sec:factorization}, we present in detail all of the serial
sparse Cholesky factorization algorithms, as implemented
in this study (MF, LL, RL, and RLB).
In Section~\ref{sec:tests}, we present timing results and storage statistics that
confirm the effectiveness of~RLB.
Section~\ref{sec:conclusion} presents our concluding remarks.

%
%

\section{Serial sparse Cholesky factorization algorithms}
\label{sec:factorization}

In the 1970s, the developers of LINPACK (later LAPACK) produced efficient software for
many problems in linear algebra, where the matrix is dense.
The key feature behind the efficiency was their development of and use of the 
Basic Linear Algebra Subroutines (BLAS),
which encapsulated computationally intensive kernels
into subroutines
that could be tuned for performance on the great variety of
computer architectures that were emerging at that time.
In particular, the level-3 BLAS often enabled excellent performance,
an example of which is matrix-matrix multiply (DGEMM).
In the 1980s, the research community was asking if
there was a way to implement sparse Cholesky factorization so that it takes
advantage of the techniques used by LAPACK for efficiency.
The breakthrough came with
the introduction of the multifrontal method in 1983 by Duff and Reid~\cite{DR83}.

Our presentation here is organized as follows.
We examine the established factorization algorithms in Section~\ref{sec:established}.
For completeness and as a point of reference in the historical development,
we present in Section~\ref{sec:gsfct}
the sparse Cholesky factorization algorithm 
used in sparse matrix packages in the 1970s.
In Section~\ref{sec:eff_background},
we present necessary notation and background material.
We then present in detail our implementation of the multifrontal method~(MF)
in Section~\ref{sec:MF}.
In Section~\ref{sec:LL}, we present in detail our implementation of the efficient
left-looking factorization algorithm~(LL) introduced in the early 1990s independently
by Rothberg and Gupta~\cite{RG91} and Ng and Peyton~\cite{NP94}.
In Section~\ref{sec:novel}, we then turn our attention to the novel variants of serial sparse Cholesky factorization
introduced in this paper.
In Section~\ref{sec:RL}, we present our new right-looking variant~RL,
and in Section~\ref{sec:RLB}, we present our new right-looking blocked
variant~RLB.

\subsection{The established algorithms}
\label{sec:established}

%
%

\subsubsection{Sparse Cholesky factorization in the 1970s}
\label{sec:gsfct}

In the 1970s, researchers introduced the first effective algorithms for sparse Cholesky
factorization.
The sparse Cholesky factorization subroutines used at that time in 
Waterloo's SPARSPAK and in
the Yale Sparse Matrix Package (YSMP)
were very similar.
In this subsection, we present the algorithm upon which they were based.

We will need the following notation.
Let~$C$ be a matrix.
We will let~$C_{*,j}$ denote column~$j$ of~$C$,
and we will let~$|C_{*,j}|$ denote the number of nonzeros in~$C_{*,j}$.
The integer vector $\glbind{j}$ contains the row indices of the nonzeros in~$L_{*,j}$
listed in ascending order.
Specifically,
\[
	\glbind{j} := \{ i : L_{i,j} \neq 0 \},
\]
with the members listed in ascending order.
The vector~$\Lnz{j}$ will
upon exit contain the nonzeros of~$L_{*,j}$, and
it will upon entry contain the corresponding entries in column~$j$ of
the lower triangle of~$A$. 
(Consequently, it is a vector
of length~$|L_{*,j}|$.)
We will use $\cdiv{j}$ to denote the action
of scaling column~$j$ to complete the factor column after all of its updates
have been received.
Finally, the algorithm will perform the computation largely within
a floating-point $n$-vector~\tn.

The algorithm is displayed in Algorithm~\ref{alg:gsfct} below.
\begin{algorithm}[htp] 
\small
\begin{flushleft}
	{\bf Input:}
  $\glbind{j}$ for every column~$j$, $1 \leq j \leq n$; \\
	For $1 \leq j \leq n$, $\Lnz{j}$ contains the appropriate entries from $A_{*,j}$. \\
	{\bf Output:}
	For $1 \leq j \leq n$, $\Lnz{j}$ contains the nonzero entries from $L_{*,j}$.
\end{flushleft}
\caption{A column-based sparse Cholesky algorithm from the 1970s.}
\label{alg:gsfct}
\begin{algorithmic}[1]
	\For{$j \gets 1$ {\bf to} $n$}
		\State{Use $\glbind{j}$ to scatter $\Lnz{j}$ (entries of $A_{*,j}$) into \tn;}
		\For{each $k < j$ {\bf such that} $j \in \glbind{k}$}
			\State{Use $\glbind{k} \cap \{j,j+1,\ldots,n\}$ to scatter-add $-L_{j:n,k} * L_{j,k}$ into \tn;}
		\EndFor
		\State{Use $\glbind{j}$ to gather $\Lnz{j}$'s modified entries back from \tn\ into $\Lnz{j}$;}
		\State{Perform $\cdiv{j}$ within $\Lnz{j}$;}
		\LineComment{$\Lnz{j}$ now contains the nonzeros of $L_{*,j}$.}
	\EndFor
\end{algorithmic}
\end{algorithm}
Since this algorithm is so simple, and moreover it is so inefficient 
on modern machines that it is no longer relevant,
we will not discuss it in detail.
We must, however, point out that the key to its efficient indexing scheme
is special structure in the Cholesky factor;
specifically, we have
\begin{equation}
	\glbind{k} \cap \{j,j+1,\ldots,n\} \subseteq \glbind{j}
	\label{eqn:inequality1}
\end{equation}
in line~4, which is the key line in the algorithm.
Note that the updates to column~$j$ are coming from the left;
hence, this is also known as a {\em left-looking} factorization algorithm.

%
%

\subsubsection{Background and notation}
\label{sec:eff_background}

As we said in Section~\ref{sec:gsfct},
we must exploit special structure in the factor matrix
to obtain an efficient indexing scheme for sparse Cholesky factorization.
Further exploitation of special structure in the factor matrix,
namely {\em supernodes}, 
opens the door to using the
level-3 BLAS in order to gain much greater efficiency.
Roughly speaking, in our context a supernode is a set of consecutive factor columns
sharing the same zero-nonzero structure below a dense lower triangular
submatrix on top,
as we saw in Figure~\ref{fig:supernode1}.
Consequently, the columns of a supernode can be organized and exploited as 
a dense submatrix of~$L$ in various sparse Cholesky factorization
algorithms.
A more formal definition of supernodes will follow our discussion
of elimination trees below.

\begin{figure}[htp]
\scriptsize
\setlength{\arraycolsep}{2pt}
\begin{center}
  \begin{tikzpicture}[scale=.75]
      \node at (-9,-.5) {\normalsize $L$};
      \node at (-9,0) [above] {
       $\left[
       \begin{array}{ccccccccc}
         {1} & & \multicolumn{7}{c}{} \\
         \ast & {2} & \multicolumn{7}{c}{} \\
         & & 3 & & \multicolumn{5}{c}{} \\
         & & \ast & 4 & \multicolumn{5}{c}{} \\
         \ast & \ast & \ast & \ast & 5 & \multicolumn{4}{c}{} \\
         \ast & + & & & \ast & 6 & \multicolumn{3}{c}{} \\
         & & \ast & + & + & \ast & 7 & \multicolumn{2}{c}{} \\
         & & \ast & \ast & \ast & + & \ast & 8 & \multicolumn{1}{c}{} \\
         \ast & \ast & & & + & \ast & + & \ast & 9
       \end{array}
	  \right]$};
%
%
      \node (1) at (-4.5,0) [circle, draw] {$1$};
      \node (2) at (-4.5,.80) [circle, draw] {$2$};
      \node (3) at (-2.5,0) [circle, draw] {$3$};
      \node (4) at (-2.5,.80) [circle, draw] {$4$};
      \node (5) at (-3.5,1.6) [circle, draw] {$5$};
      \node (6) at (-3.5,2.40) [circle, draw] {$6$};
      \node (7) at (-3.5,3.2) [circle, draw] {$7$};
      \node (8) at (-3.5,4.0) [circle, draw] {$8$};
      \node (9) at (-3.5,4.8) [circle, draw] {$9$};
%
      \draw (1) -- (2);
      \draw (2) -- (5);
      \draw (3) -- (4);
      \draw (4) -- (5);
      \draw (5) -- (6);
      \draw (6) -- (7);
      \draw (7) -- (8);
      \draw (8) -- (9);
  \end{tikzpicture}
  \caption{The sparse Cholesky factor shown in Figure~\ref{fig:supernode1},
  along with its elimination tree.}
  \label{fig:factor}
\end{center}
\end{figure}
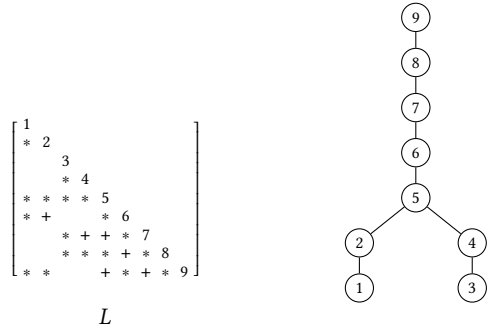

The {\em elimination tree} (or forest) associated with a sparse Cholesky factor
has proved to be extremely valuable in sparse matrix computations
(see Liu~\cite{Liu90} for a survey).
For each column~$j$, the parent of~$j$
in the elimination tree is defined by
\[
	p(j) := \min \{ i : i \in \glbind{j} \setminus \{j\} \};
\]
if $\glbind{j} = \{j\}$, then $p(j) := \nul$,
and $j$ is a root.
A well known property of elimination trees~\cite{Liu90,Schreiber82} is
\[
	\glbind{j} \setminus \{j\} \subseteq \glbind{p(j)},
\]
which can be viewed as a special case of set containment~(\ref{eqn:inequality1})
in Section~\ref{sec:gsfct}.
Figure~\ref{fig:factor} displays the elimination tree associated
with the Cholesky factor in Figure~\ref{fig:supernode1}.

We are now in a position to define supernodes;
we will first limit our attention to the {\em fundamental supernode partition}.
It is defined as follows:
columns~$j$ and~$p(j)$ belong to the the same supernode if
and only if
\[
	\glbind{j} \setminus \{j\} = \glbind{p(j)}
\]
and moreover~$j$ is the only child of~$p(j)$ in the elimination tree.
Liu, Ng, and Peyton~\cite{LNP93} introduced an efficient 
algorithm for computing the fundamental supernode partition
during the symbolic factorization step.
The reader may verify from the elimination tree shown
in Figure~\ref{fig:factor} that the supernode partition
in Figure~\ref{fig:supernode1} is indeed the fundamental supernode partition.
In order to improve factorization times further,
we will later coarsen the supernode partition (see Ashcraft and Grimes~\cite{AG89}).

The following notation is relevant to all of the supernodal
sparse Cholesky algorithms discussed in our paper.
We will let~$L_{*,J}$ denote the submatrix of~$L$ comprising
all of the columns in supernode~$J$,
and we will let~$A_{*,J}$ be the corresponding columns of~$A$.
Let~$f$ be the first column index of supernode~$J$, and let~$\ell$ be the last
column index of supernode~$J$.
In the algorithms, $\Lnz{J}$ will be a 
$|L_{*,f}|$ by $|J|$
block of storage that upon exit
contains the nonzeros of~$L_{*,J}$ and upon entry contains the corresponding
entries in the lower triangle of~$A$.
We will define
\[
  \glbind{J} := \glbind{f}.
\]
Observe in Figure~\ref{fig:supernode1} that
\[
  \glbind{J_1}  = \left[ \begin{array}{c}
                           1 \\ 2 \\ 5 \\ 6 \\ 9
                         \end{array}
                  \right]
  \mbox{and }
  \glbind{J_2}  = \left[ \begin{array}{c}
                           3 \\ 4 \\ 5 \\ 7 \\ 8
                         \end{array}
                  \right].
\]
We will define the parent function of the
{\em supernodal elimination tree} by $p(J) := J'$
if $k \in J'$, where
\[
  k = \min \{ i : i \in \glbind{J} \mbox{ and } i > \ell \};
\]
if $\glbind{J} = J$, then $p(J) := \nul$, and~$J$ is a root.
Observe in Figure~\ref{fig:supernode1} that 
$p(J_1) = J_3$, 
$p(J_2) = J_3$, and
$p(J_3) = \nul$.

In the algorithms, we will use $\cdiv{J}$ to denote the action
of completing all of the columns of~$J$ after all of~$J$'s 
updates have been received.
In practice, this is done by performing a dense Cholesky factorization
on the lower triangular matrix atop supernode~$J$ (using DPOTRF),
followed by a block triangular solve (using DTRSM) to obtain
the part of the supernode below the triangle.

Finally, we will need indices that replace each global index with its location
in a given ancestor's index list.
Such {\em relative indices} were first introduced by Schreiber~\cite{Schreiber82}
and were later used by Ashcraft~\cite{Ashcraft87}
to improve the multifrontal method.
(In their case, the given ancestor was always the parent.)
In the set of relative indices $\relind{J,J'}$,
where~$J'$ is an ancestor of~$J$ in the supernodal elimination tree,
each global index~$i \in \glbind{J} \cap \glbind{J'}$ is replaced
in $\relind{J,J'}$ by the {\em distance} of~$i$ from the bottom
of the list $\glbind{J'}$.
For example, in Figure~\ref{fig:supernode1} we have
\[
  \relind{J_1,J_3}  = \left[ \begin{array}{c}
                               4 \\ 3 \\ 0
                             \end{array}
                      \right]
  \mbox{and }
  \relind{J_2,J_3}  = \left[ \begin{array}{c}
                               4 \\ 2 \\ 1 \\
                             \end{array}
                      \right].
\]
Such indices are vital to the assembly operations
alluded to in Section~\ref{sec:intro}.


%
%

\subsubsection{The multifrontal method (MF)}
\label{sec:MF}

During the multifrontal method,
the supernodes are processed in a postordering of the supernodal
elimination tree.
It is well known that~MF manages a stack of update matrices throughout
the computation.
The amount of floating-point storage needed for this stack depends on
the order of the siblings in the supernodal elimination tree
used to determine the postordering.
Liu~\cite{Liu86b} introduced an algorithm that orders the siblings
within the supernodal elimination tree in such a way that
minimizes the storage for this stack over all such sibling reorderings.
We will use this technique from Liu~\cite{Liu86b} to reduce
stack storage size in all of our~MF runs.

A detailed presentation of our implementation of the multifrontal method is
shown in Algorithm~\ref{alg:MF}.
\begin{algorithm}[htp] 
\small
\begin{flushleft}
	{\bf Input:}
	$p(J)$ and $\glbind{J}$ for every supernode~$J$; \\
	for every supernode~$J$, $\Lnz{J}$ contains the appropriate entries from $A_{*,J}$. \\
	{\bf Output:}
	For every supernode~$J$, $\Lnz{J}$ contains the nonzeros of $L_{*,J}$.
\end{flushleft}
\caption{Our implementation of multifrontal Cholesky factorization (MF).}
\label{alg:MF}
	\begin{algorithmic}[1]
	\LineComment{Initializations}
	\State{Transform every list $\glbind{J} \cap \glbind{p(J)}$ into $\relind{J,p(J)}$, provided $p(J)$ exists;}
	\State \multiline{Initialize to zero the storage for the stack of update matrices, 
	and initialize the stack as empty;}
	\For{each supernode $J$ (in postorder)}
		\State{$f \gets \min\{i:i \in J\}$; $\ell \gets \max\{i:i \in J\}$;}
		\LineComment{Assemble the update matrices of $J$'s children into $J$'s update matrix.}
		\For{each child $C$ of $J$ (popped from the stack)}
			\If{$C$ is the first child popped from the stack}
				\State{Pop $U_C^p$ from the stack and use $\relind{C,J}$ to {\em extend} $U_C^p$ into $U_J^s$ (in place);}
			\Else
				\State{Pop $U_C^p$ from the stack and use $\relind{C,J}$ to {\em assemble} $U_C^p$ into $U_J^s$;}
			\EndIf
		\EndFor
		\State{Perform $\cdiv{J}$ within $\Lnz{J}$ (DPOTRF, DTRSM);}
		\LineComment{$\Lnz{J}$ now contains the nonzeros of $L_{*,J}$.}
		\State{Compute and accumulate in $U_J^s$ the lower triangle of $-L_{\ell+1:n,J} * L_{\ell+1:n,J}^T$ (DSYRK);}
		\LineComment{Update $J$'s parent.}
		\If{$p(J)$ exists}
			\State{Use $\relind{J,p(J)}$ to assemble $p(J)$'s columns within $U_J^s$ into $\Lnz{p(J)}$;}
			\State{Remove the columns assembled above (in line~15) from $U_J^s$;}
		\EndIf
		\IIf{$U_J^s \neq \nul$} Pack $U_J^s$ into $U_J^p$ as $U_J^p$ is pushed onto the stack; \EndIIf
	\EndFor
	\State{Transform each list $\relind{J,p(J)}$ into $\glbind{J} \cap \glbind{p(J)}$, provided $p(J)$ exists;}
\end{algorithmic}
\end{algorithm}
For a more thorough look at the multifrontal method,
consult Liu's review article~\cite{Liu92}.

Input into the algorithm are the supernodal elimination tree,
the global indices for each supernode,
and the lower triangle of~$A$ stored within the data structure
that will ultimately hold the nonzeros of~$L$.
In line~1, the global indices for each supernode~$J$
are replaced by the corresponding indices relative to the parent~$p(J)$.
In line~2, all of the storage for the stack of update matrices is initialized
with zero entries, and the stack is initialized to be an empty stack.
Lines~3--19 constitute the main loop through the supernodes
in postorder.

Line~4 initializes~$f$ and~$\ell$ to the first and last column
indices of supernode~$J$, respectively.
Lines~5--11 assemble update matrices from the children~$C$
of~$J$ into~$J$'s update matrix, which is initially zero.
First, we need to explain some notation.
An update matrix~$U_C^p$ popped from the stack is a dense lower
triangular matrix stored in {\em packed} form,
which means that it stores only the nonzeros of the matrix.
It contains all of the needed updates from~$C$ and~$C$'s
descendants in the supernodal elimination tree.
The update matrix~$U_J^s$ currently being created will
ultimately be a dense lower triangular matrix stored as a
square two-dimensional array.

Lines~5--11 process every child~$C$ of~$J$ that has an update
matrix on the stack.
For the first child~$C$ popped from the stack,
the  update matrix~$U_C^p$ is {\em extended} (i.e., scattered)
into~$U_J^s$ in line~7.
This is done {\em in place} so the storage for the two overlaps.
For each subsequent child~$C'$ popped from the stack,
the  update matrix~$U_{C'}^p$ is {\em assembled} (i.e., scatter-added)
into~$U_J^s$ in line~9.

When~MF gets to line~12,
all of the updates to supernode~$J$ from descendant supernodes
have been received.
Line~12 uses DPOTRF and DTRSM to complete the 
computation of~$L_{*,J}$ within $\Lnz{J}$.
Line~13 then uses DSYRK to accumulate all of supernode~$J$'s updates 
for its ancestors into~$U_J^s$.

Next, if~$J$ has a parent~$p(J)$, then line~15 uses $\relind{J,p(J)}$
to assemble~$p(J)$'s columns within~$U_J^s$ into $\Lnz{p(J)}$.
In line~16, the columns assembled in line~15 are removed from~$U_J^s$.
In line~18, if~$U_J^s$ still has active columns, then
the active portion of~$U_J^s$ is packed as $U_J^p$ as it is
pushed onto the top of the stack.

In line~20, the relative indices computed in line~1
are transformed back into global indices.
This is important because our version of the triangular solves
used in step~4 of the solution process requires the global indices.

%
%

\subsubsection{Left-looking supernodal sparse factorization (LL)}
\label{sec:LL}

The left-looking supernodal sparse Cholesky factorization algorithm
can probably best be viewed as a block version of the left-looking
factorization algorithm from the~1970s discussed in Section~\ref{sec:gsfct}.
A detailed presentation of our implementation of this algorithm is
shown in Algorithm~\ref{alg:blkfct}.
\begin{algorithm}[htp] 
\small
\begin{flushleft}
	{\bf Input:}
	$\glbind{J}$ for every supernode~$J$; \\
	for every supernode~$J$, $\Lnz{J}$ contains the appropriate entries from $A_{*,J}$. \\
	{\bf Output:}
	For every supernode~$J$, $\Lnz{J}$ contains the nonzeros of $L_{*,J}$.
\end{flushleft}
\caption{Our detailed left-looking supernodal Cholesky algorithm (LL).}
\label{alg:blkfct}
	\begin{algorithmic}[1]
	\LineComment{Initializations}
	\State{Initialize to zero the storage for the update matrices (i.e., the $U_{K,J}^s$'s);}
	\For{each supernode $J$ (in ascending order)}
		\State{$f \gets \min \{ i : i \in J\}$; $\ell \gets \max \{ i : i \in J\}$;}
		\State{Use $\glbind{J}$ to scatter $J$'s relative indices into \indmapn;}
		\For{each supernode $K$ {\bf such that} $\glbind{K} \cap J \neq \emptyset$}
			\If{$|K|=1$}
				\State{Let $k$ be the sole member of $K$;}
				\State \multiline{%
					Compute the lower trapezoid of $-L_{f:n,k} * L_{f:\ell,k}^T$,
					and use $\glbind{K} \cap \{f,f+1,\ldots,n\}$
					and \indmapn\ to incorporate these updates directly into $\Lnz{J}$;}
			\Else
				\If{$K$'s update matrix for~$J$ is dense with respect to its target in $\Lnz{J}$}
					\State \multiline{%
						Compute the lower trapezoid of $-L_{f:n,K} * L_{f:\ell,K}^T$,
						and incorporate these updates directly into $\Lnz{J}$ (DSYRK, DGEMM);}
				\Else
					\State \multiline{%
						Compute the lower trapezoid of $-L_{f:n,K} * L_{f:\ell,K}^T$
						within $U_{K,J}^s$ (DSYRK, DGEMM);}
					\State \multiline{%
						Use $\glbind{K} \cap \{f,f+1,\ldots,n\}$ to gather
						$\relind{K,J}$ from \indmapn;}
					\State \multiline{%
						Use $\relind{K,J}$ to assemble $U_{K,J}^s$
						into $\Lnz{J}$;}
				\EndIf
			\EndIf
		\EndFor
		\State{Perform $\cdiv{J}$ within $\Lnz{J}$ (DPOTRF, DTRSM);}
		\LineComment{$\Lnz{J}$ now contains the nonzeros of $L_{*,J}$.}
	\EndFor
\end{algorithmic}
\end{algorithm}
The input into~LL is the same as the input into~MF,
except that~LL has no need for the supernodal elimination tree.

In line~1, every entry in the block of floating-point storage used
for the update matrices is initialized to zero.
This block of working storage is allocated before the factorization
so that it is just large enough to handle the largest update matrix
that will be stored during the course of the computation.
To clarify notation here,
we will let~$U_{K,J}^s$ denote the dense lower trapezoidal update
matrix that supernode~$K$ contributes to supernode~$J$, stored
as a two-dimensional array.

Lines~2--20 constitute the main loop through the supernodes
in ascending order.
Line~3 initializes~$f$ and~$\ell$ to the first and last column indices of
supernode~$J$, respectively.
In line~4, supernode~$J$'s global indices are {\em translated}
into local indices and scattered into an integer vector \indmapn\ for
use later in line~14, as we shall see.
As an example, for supernode~$J_3$ in Figure~\ref{fig:supernode1},
we would have
$\indmap{5} = 4$,
$\indmap{6} = 3$,
$\indmap{7} = 2$,
$\indmap{8} = 1$, and
$\indmap{9} = 0$
after line~4 is executed.

Lines~5--18 constitute the loop that updates supernode~$J$ with
supernodes to the left of~$J$.
Lines~7--8 comprise special code to take care of supernodes~$K$
having only one column.
We omit discussing the details here, except to say that the update
from the single column~$k$ of~$K$ is incorporated {\em directly} into
factor storage within~$\Lnz{J}$ in line~8;
that is, no update matrix~$U_{K,J}^s$ is stored.

In lines~10--16, supernode~$K$ has two or more columns.
Line~11 uses BLAS routines DSYRK and DGEMM to incorporate
the dense lower trapezoidal update matrix from supernode~$K$ {\em directly} into
two dense submatrices within $\Lnz{J}$.
As before,
no update matrix~$U_{K,J}^s$ is stored;
it is not necessary in this case.

Lines~13--15 are the key lines in the algorithm:
in practice, most of the updates from one supernode
to another are performed by these lines.
Here, it is the case that the update matrix from supernode~$K$
is sparse relative to its target within $\Lnz{J}$.
Line~13 uses DSYRK and DGEMM to compute the dense lower
trapezoidal update matrix~$U_{K,J}^s$, 
stored as a two-dimensional array.

Line~14 uses the global indices of~$K$ to gather $\relind{K,J}$
from \indmapn.
For example, consider the action taken by the algorithm when it
executes line~14 as it uses supernode~$J_1$ to update
supernode~$J_3$ in Figure~\ref{fig:supernode1}.
It then uses 
\[
  \glbind{J_1} \cap \{5,6,7,8,9\} = \left[\begin{array}{c} 5 \\ 6 \\ 9 \end{array}\right]
\]
to gather
\[
  \relind{J_1,J_3} = \left[\begin{array}{c} 4 \\ 3 \\ 0 \end{array}\right]
\]
from the current contents of \indmapn.
Note that $\relind{K,J}$ here is stored in an integer
work vector.
Line~15 completes this process by using $\relind{K,J}$ to assemble
$U_{K,J}^s$ into factor storage within $\Lnz{J}$.

Upon exit from the loop in lines~5--18, supernode~$J$ has received all of its updates from 
supernodes to its left.
Line~19 uses DPOTRF and DTRSM to complete the columns of~$L_{*,J}$ within $\Lnz{J}$.

\subsection{Two novel factorization variants}
\label{sec:novel}

%
%

\subsubsection{A right-looking method (RL)}
\label{sec:RL}

As far as we know, the factorization method presented in this subsection
has not yet appeared in the literature.
The algorithm presented here is somewhat similar to the
multifrontal method, but it is significantly
simpler, in our view.
A detailed presentation of our implementation of an efficient 
right-looking supernodal sparse
Cholesky factorization algorithm~(RL) is shown in Algorithm~\ref{alg:RL}.
\begin{algorithm}[htp] 
\small
\begin{flushleft}
	{\bf Input:}
	$p(J)$ and $\glbind{J}$ for every supernode~$J$; \\
	for every supernode~$J$, $\Lnz{J}$ contains the appropriate entries from $A_{*,J}$. \\
	{\bf Output:}
	For every supernode~$J$, $\Lnz{J}$ contains the nonzeros of $L_{*,J}$.
\end{flushleft}
\caption{Our detailed right-looking supernodal Cholesky algorithm (RL).}
\label{alg:RL}
	\begin{algorithmic}[1]
	\LineComment{Initializations}
	\State{Transform every list $\glbind{J} \cap \glbind{p(J)}$ into $\relind{J,p(J)}$, provided $p(J)$ exists;}
	\State{Initialize to zero the storage for the update matrices (i.e., the $U_{J}^s$'s);}
	\For{each supernode $J$ (in ascending order)}
		\State{$f \gets \min\{i:i \in J\}$; $\ell \gets \max\{i:i \in J\}$;}
		\State{Perform $\cdiv{J}$ within $\Lnz{J}$ (DPOTRF, DTRSM);}
		\LineComment{$\Lnz{J}$ now contains the nonzeros of $L_{*,J}$.}
		\State{Compute and accumulate in $U_J^s$ the lower triangle of $-L_{\ell+1:n,J} * L_{\ell+1:n,J}^T$ (DSYRK);}
		\LineComment{Update $J$'s ancestors.}
		\State{$P \gets p(J)$;}
		\While{$U_J^s \neq \nul$}
			\If{$P > p(J)$}
				\State{$\relind{J,P} \gets \relind{J,C} \circ \relind{C,P}$;}
			\EndIf
			\If{$U_J^s$ has columns from $P$}
				\State{Use $\relind{J,P}$ to assemble $U_J^s$'s columns from $P$ into $\Lnz{P}$;}
				\State{Remove the columns assembled above (in line~13) from $U_J^s$;}
			\EndIf
			\State{$C \gets P$; $P \gets p(P)$;}
			\Comment{Next ancestor $P$}
		\EndWhile
	\EndFor
	\State{Transform each list $\relind{J,p(J)}$ into $\glbind{J} \cap \glbind{p(J)}$, provided $p(J)$ exists;}
\end{algorithmic}
\end{algorithm}

The input to and the output from~RL and~MF
are the same.
In line~1, the global indices for each supernode~$J$
are replaced by the corresponding indices relative to the parent~$p(J)$.
In line~2, every entry in the block of floating-point storage used
for the update matrices is initialized to zero.
This block of working storage is allocated before the factorization
so that it is just large enough to handle the largest update matrix
that will be stored during the course of the computation.
To clarify notation here,
we will let~$U_{J}^s$ denote the dense lower triangular update
matrix that supernode~$J$ contributes to its ancestor supernodes,
stored as a two-dimensional square array.
Observe that there is no stack of update matrices to manage, and no
associated issues of limiting the size of stack storage;
we view this as a significant simplification over what is
required by~MF.

Lines~3--18 constitute the main loop through the supernodes
in ascending order.
(Recall that~MF requires a postordering of the supernodes.)
Line~4 initializes~$f$ and~$\ell$ to the first and last column
indices of supernode~$J$, respectively.
When~RL gets to line~5,
all of the updates to supernode~$J$ from supernodes to its left
have been received.
Line~5 uses DPOTRF and DTRSM to complete the 
computation of~$L_{*,J}$ within $\Lnz{J}$.
Line~6 then uses DSYRK to accumulate all of supernode~$J$'s updates 
for its ancestors into~$U_J^s$.

In lines~7--17, for each ancestor supernode~$P$ of~$J$ updated by~$J$,
those needed updates are assembled into factor storage $\Lnz{P}$.
This loop is the key to the algorithm.
Starting at $P=p(J)$ (see line~7),
this {\bf while} loop proceeds up the supernodal elimination tree,
updating ancestors of~$J$ as needed.
Line~10 is the key line in the {\bf while} loop.
Line~10 will be executed for any ancestor~$P$ of~$J$ above~$p(J)$.
When line~10 is executed,
$\relind{J,C}$, where~$C$ is the child of~$P$
between~$J$ and~$P$, has already been computed and stored.
The indices $\relind{C,P}$ are also available;
they were computed in line~1 of the algorithm.

The example that we have been using throughout the paper is
too simple to help us illustrate what is going on during
the execution of line~10.
So, we will create a hypothetical pair of index vectors
to serve as our illustration.
Suppose that we have
\[
  \relind{J,C}  = \left[ \begin{array}{c}
                               5 \\ 3 \\ 1
                             \end{array}
                      \right]
  \mbox{and }
  \relind{C,P}  = \left[ \begin{array}{c}
                               \ast \\ \ast \\ 7 \\ \ast \\ 4 \\ \ast \\ 2 \\ \ast
                             \end{array}
                      \right]
\]
immediately before line~10 is executed.
When $\relind{J,C} \circ \relind{C,P}$ is computed,
the index~1 in $\relind{J,C}$ provides the {\em distance}
of its replacement in $\relind{C,P}$ from the 
bottom of $\relind{C,P}$;
thus, its replacement is~2.
And so we see that the indices~3 and~5 in $\relind{J,C}$
will be replaced by~4 and~7 in $\relind{C,P}$, respectively.
To summarize,
\[
  \relind{J,P}  = \left[ \begin{array}{c}
                               7 \\ 4 \\ 2
                             \end{array}
                      \right]
\]
after line~10 has been executed.
All of this can be effected with a simple gather operation.

In line~12, the first index in $\relind{J,P}$
is used to determine if~$P$ has any columns belonging
to~$U_J^s$.
If it does, then
line~13 uses $\relind{J,P}$
to assemble~$P$'s columns within~$U_J^s$ into $\Lnz{P}$.
In line~14, the columns assembled in line~13 are 
removed from~$U_J^s$.
Line~16 proceeds up the tree to the next ancestor~$p(P)$.

In line~19, after the computation has been completed,
the relative indices computed in line~1 are transformed back into global indices.

%
%

\subsubsection{A right-looking blocked method (RLB)}
\label{sec:RLB}

A detailed presentation of our implementation of efficient right-looking blocked supernodal
sparse Cholesky factorization (RLB) is shown in Algorithm~\ref{alg:RLB}.
\begin{algorithm}[htp] 
\small
\begin{flushleft}
	{\bf Input:}
	$p(J)$ and $\glbind{J}$ for every supernode~$J$; \\
	The set of blocks for each supernode $J$; \\
	for every supernode~$J$, $\Lnz{J}$ contains the appropriate entries from $A_{*,J}$. \\
	{\bf Output:}
	For every supernode~$J$, $\Lnz{J}$ contains the nonzeros of $L_{*,J}$.
\end{flushleft}
\caption{Our right-looking blocked supernodal Cholesky algorithm (RLB).}
\label{alg:RLB}
	\begin{algorithmic}[1]
	\LineComment{Initializations}
	\State{Transform every list $\glbind{J} \cap \glbind{p(J)}$ into $\relind{J,p(J)}$, provided $p(J)$ exists;}
	\For{each supernode $J$ (in ascending order)}
		\State{$f \gets \min\{i:i \in J\}$; $\ell \gets \max\{i:i \in J\}$;}
		\State{Perform $\cdiv{J}$ within $\Lnz{J}$ (DPOTRF, DTRSM);}
		\LineComment{$\Lnz{J}$ now contains the nonzeros of $L_{*,J}$.}
					\State{Extract $\relindB{J,p(J)}$ from $\relind{J,p(J)}$;}
		\LineComment{Update $J$'s ancestors.}
		\State{$P \gets p(J)$;}
		\While{there remains an ancestor supernode of $J$ to update}
			\If{$P > p(J)$}
				\State{$\relindB{J,P} \gets \relindB{J,C} \circ \relind{C,P}$;}
			\EndIf
			\For{each block $B$ of $J$ {\bf such that} $B \subseteq P$ (in order)}
				\State \multiline{%
					Using the entry in $\relindB{J,P}$ for $B$, perform the lower triangle of the update \\
					$L_{B,B} \gets L_{B,B} - L_{B,J} * L_{B,J}^T$ directly into $\Lnz{P}$ (DSYRK);}
					\For{each ``maximal'' block $B'$ of $J$ below $B$ (in order)}
						\State \multiline{%
							Using the entries in $\relindB{J,P}$ for $B$ and $B'$, perform the update \\
							$L_{B',B} \gets L_{B',B} - L_{B',J} * L_{B,J}^T$ directly into $\Lnz{P}$ (DGEMM);}
					\EndFor
			\EndFor
			\State{$C \gets P$; $P \gets p(P)$;}
			\Comment{Next ancestor $P$}
		\EndWhile
	\EndFor
	\State{Transform each list $\relind{J,p(J)}$ into $\glbind{J} \cap \glbind{p(J)}$, provided $p(J)$ exists;}
\end{algorithmic}
\end{algorithm}
As we said in Section~\ref{sec:intro},~RLB can be viewed as a modification 
of~RL.
When~RL is processing supernode~$J$, it uses one call to DSYRK to compute
all of~$J$'s updates to its ancestors
within an update matrix,
and then it assembles these updates into the appropriate
target ancestor supernodes, one by one.
By contrast, when~RLB is processing supernode~$J$, it decomposes the updating process
into many calls to DSYRK and DGEMM, where each is updating factor
matrix storage directly;
consequently, RLB stores and assembles no update matrices.

The input to~RLB matches that of~RL, except there is one additional item.
For each supernode~$J$, we will need its {\em blocks}.
These are stored as a list of block sizes, where the list is in
the order of the blocks of~$J$ from top to bottom in the supernode.

The first four lines in~RLB also begin as in~RL, except there is one line from~RL
that is missing;
there is no line to zero out space for update matrices
since there are none used in the algorithm.

The indexing in~RLB is performed the same way it is in~RL,
except that far fewer indices are involved, in practice.
In~RLB, there is a single index for each {\em block} in the supernode
rather than a single index for each {\em row} in the supernode.
The needed indices are extracted in line~5.
As an example, consider the indices of~$J_1$ and~$J_2$
relative to~$J_3$ in Figure~\ref{fig:supernode1}.
(Recall that $p(J_1) = J_3$ and $p(J_2) = J_3$.)
We have
\[
  \relind{J_1,J_3}  = \left[ \begin{array}{c}
                               4 \\ 3 \\ 0
                             \end{array}
                      \right]
  \mbox{and }
  \relind{J_2,J_3}  = \left[ \begin{array}{c}
                               4 \\ 2 \\ 1
                             \end{array}
                      \right].
\]
Immediately after the execution of line~5, for~$J_1$ and later for $J_2$, we have
\[
  \relindB{J_1,J_3}  = \left[ \begin{array}{c}
                                4 \\ 0
                              \end{array}
                       \right]
  \mbox{and }
  \relindB{J_2,J_3}  = \left[ \begin{array}{c}
                                4 \\ 2 
                              \end{array}
                       \right],
\]
respectively.

Lines~11--16 are the lines that sharply distinguish this algorithm
from the others.
The two nested {\bf for} loops process every distinct
pair of blocks~$B$ and~$B'$ in supernode~$J$,
such that $B \subseteq P$ and
$B'=B$ or~$B'$ is below~$B$.
Line~12 takes care of the case where block~$B$ is paired with itself.
Here, a DSYRK operation is performed,
and the target of the update is in factor storage; 
more precisely it is in $\Lnz{P}$.
In line~12, a single index from $\relindB{J,P}$
is used to produce the pointer to the target in $\Lnz{P}$.
Line~14 takes care of the case where block~$B$ is paired with a different block~$B'$.
Here, a DGEMM operation is performed,
and the target of the update is again in $\Lnz{P}$.
In line~14, a pair of indices from $\relindB{J,P}$
are used to produce the pointer to the target in $\Lnz{P}$.
Upon exit from the nested {\bf for} loops,
all updates from supernode~$J$ have been incorporated
into supernode~$P$.

There is one addition detail concerning line~13.
The word ``maximal'' is used there to indicate that,
whenever possible, contiguous blocks are merged together
across supernode boundaries to form a larger block~$B'$.
We have tested with and without this feature,
and we have found that this feature consistently results
in slightly better factorization times.

In line~20, after the computation has been completed,
the relative indices computed in line~1 are transformed back into global indices.

%
%

\section{Testing the factorization methods}
\label{sec:tests}

For this section, we ran some tests to compare the factorization methods
described in Section~\ref{sec:factorization}.
Section~\ref{sec:proceedure} details how this testing was carried out,
and Section~\ref{sec:results} describes the results of the testing.

%
%

\subsection{How the testing was carried out}
\label{sec:proceedure}

For our testing, we selected a set of matrices from the
SuiteSparse collection~\cite{DH11}
of sparse matrices.
We included every symmetric matrix for which $n \geq$ 500,000,
with the following restrictions.
We included only matrices that could be in a realistic sparse
linear system to solve.
Specifically, we excluded graphs from social networks and all other
graphs in no way connected to a linear system.
There is one inconsistency in our selection criteria,
due to practical considerations.
We included the matrices nlpkkt80 and nlpkkt120 (two related optimization matrices),
but we excluded the other matrices in this particular family of matrices
because they require too much storage and time to factor.
Ultimately, we included~36 matrices in our testing.

In the ordering step of the solution process, 
we use the nested dissection routine from {\sc metis}~\cite{KK99}
to generate the fill-reducing ordering.
This is followed by the symbolic factorization step;
we need to discuss in some detail two tasks that are
performed during this step.

First, we need to discuss how supernodes are produced.
Within the symbolic factorization step,
the fundamental supernode partition is computed first.
Most often, however, this supernode partition will be 
so fine near the bottom of the tree that it hinders
good factorization performance.
Ashcraft and Grimes~\cite{AG89} introduced the idea of 
merging supernodes together,
thereby coarsening the supernode partition in order
to improve factorization performance.
This has become a standard practice in software for
sparse symmetric factorization.
For example, both the MA87 package~\cite{HRS10}
and the MA57 package~\cite{Duff04} perform such supernode merging by default.
In our symbolic factorization step, we will do likewise;
the details follow.

Our supernode-merging algorithm merges a sequence of child-parent
pairs~$J$ and~$p(J)$ until a stopping criterion is satisfied.
As the next pair~$J$ and~$p(J)$ to merge into a single supernode,
the algorithm chooses a pair whose merging creates the minimum
amount of new fill in the factor matrix.
(To do this requires the use of a heap.)
The merging stops whenever the next merging operation will cause
the cumulative percentage increase in factor storage to 
exceed~$12.5$ percent.
(We have tested various values for the percentage,
and~$12.5$ percent works well.)
For every test matrix, the associated increase in factorization 
work never exceeds one percent.
By doing it in this way,
we are trying to normalize supernode merging across
our set of test matrices.

During the symbolic factorization step,
after the supernode partition has been coarsened,
we reorder the columns within supernodes using the
{\em partition refinement} (PR) method~\cite{JNP18,KNP24}.
It is essential to do so for the factorization
method~RLB,
since RLB is not at all a viable method when there is no
reordering of columns within supernodes.
(The~TSP reordering method would also work equally well,
but at a greater cost to compute~\cite{KNP24}.)
We also compute the~PR reordering for the other
factorization methods (MF,~LL, and~RL).
For these methods, the~PR reordering results in modest factorization
time reductions; nonetheless, these reductions more that offset the cost in time of computing the~PR
reorderings, which is very small.

We ran the experiments on a dual socket machine containing two Cascade Lake 
Intel(R) Xeon(R) Gold~5220R processors (2.20GHz cpus) 
with~24 cores per socket (48~cores total) and 256~GB of memory.
The 256~GB of memory is split evenly across two NUMA domains, with one NUMA domain per socket.
We compiled our Fortran code with the gfortran compiler
using the optimization flag~-O3.
We performed runs where Intel's MKL serial BLAS were linked in, and
we performed runs where Intel's MKL multithreaded BLAS were linked in.
In the latter case, we ran the experiments with OpenMP affinity enabled
because this improved performance somewhat.

%
%

\subsection{Results}
\label{sec:results}

Consider a list of our~36 matrices,
where the matrices are listed in ascending order by the
number of floating-point operations required by numerical
factorization.
Each of the first fifteen matrices in the list requires
less than ten seconds to factor using any of our four
factorization methods, when the serial BLAS are linked in.
We will refer to these matrices as our {\em small} matrices;
they are ``small'' in the sense that the nested dissection
ordering is able to preserve a great deal of the sparsity
of~$A$ in the factor matrix~$L$.
We feel that it is important to include such matrices
in our study.
The factorization times (in seconds) for these fifteen matrices
are displayed in Table~\ref{tab:small}.

\begin{table}[htb]
\begin{center}
\begin{tabular}{|l||r|r|r|r|} \hline
\multicolumn{1}{|c||}{Matrices} &
\multicolumn{1}{c|}{MF} &
\multicolumn{1}{c|}{LL} &
\multicolumn{1}{c|}{RL} &
\multicolumn{1}{c|}{RLB} \\ \hline
lp1            &  0.172  &   \bf{0.070} &   0.171  &   0.170   \\
bundle\_adj     &  0.254  &   0.241  &   0.228  &   \bf{0.209}   \\
parabolic\_fem  &  \bf{0.385} &   0.483  &   0.388  &   0.476   \\
tmt\_sym        &  0.520  &   0.655  &   \bf{0.519} &   0.660   \\
ecology1       &  \bf{0.711} &   0.870  &   0.714  &   1.002   \\
thermal2       &  0.843  &   1.079  &   \bf{0.842} &   1.094   \\
G3\_circuit     &  \bf{1.818} &   2.174  &   1.823  &   2.379   \\
af\_shell1      &  1.501  &   1.515  &   1.488  &   \bf{1.353}   \\
af\_0\_k101      &  1.696  &   1.716  &   1.677  &   \bf{1.530}   \\
gsm\_106857     &  2.066  &   2.101  &   2.061  &   \bf{1.961}   \\
ldoor          &  2.219  &   2.250  &   2.200  &   \bf{2.001}   \\
inline\_1       &  3.521  &   3.508  &   3.463  &   \bf{3.183}   \\
apache2        &  4.165  &   4.198  &   \bf{4.063} &   4.108   \\
boneS10        &  6.844  &   6.796  &   6.706  &   \bf{6.237}   \\
af\_shell10     &  9.219  &   9.240  &   9.090  &   \bf{8.480}   \\ \hline
\end{tabular}
\caption{Factorization times (in seconds) for the fifteen small matrices
whenever the serial BLAS are linked in.
The fastest times are in bold typeface.}
\label{tab:small}
\end{center}
\end{table}

The fastest times are displayed in bold typeface.
For each of the top six matrices in the table,
the fastest time is less than one second.
RLB is the fastest method for only one of these six matrices.
For each of the bottom nine matrices in the table,
the fastest time is greater than one second.
RLB is the fastest method for seven of these nine matrices.
We see that~RLB is unequivocally the fastest of the four methods
overall for this set of small matrices.
With such small factorization times involved,
we chose not to investigate the use of multithreaded BLAS
on these matrices.

Consider the remaining~21 matrices at the bottom of our list of matrices in ascending
order by factorization operations count;
we will refer to these matrices as our {\em large} matrices.
The number of factorization operations for these matrices
varies widely from a low of $8.99 \times 10^{11}$ for matrix Curl\_Curl\_2 to a high
of $2.64 \times 10^{14}$ for matrix Queen\_4147.
We will use performance profiles to compare the four factorization algorithms
on these large matrices.

\begin{figure}[htb]
\begin{center}
  \caption{Performance profile for the factorization times for the~21 large matrices
           whenever the serial BLAS are linked in.}
  \includegraphics[width=.8\textwidth]{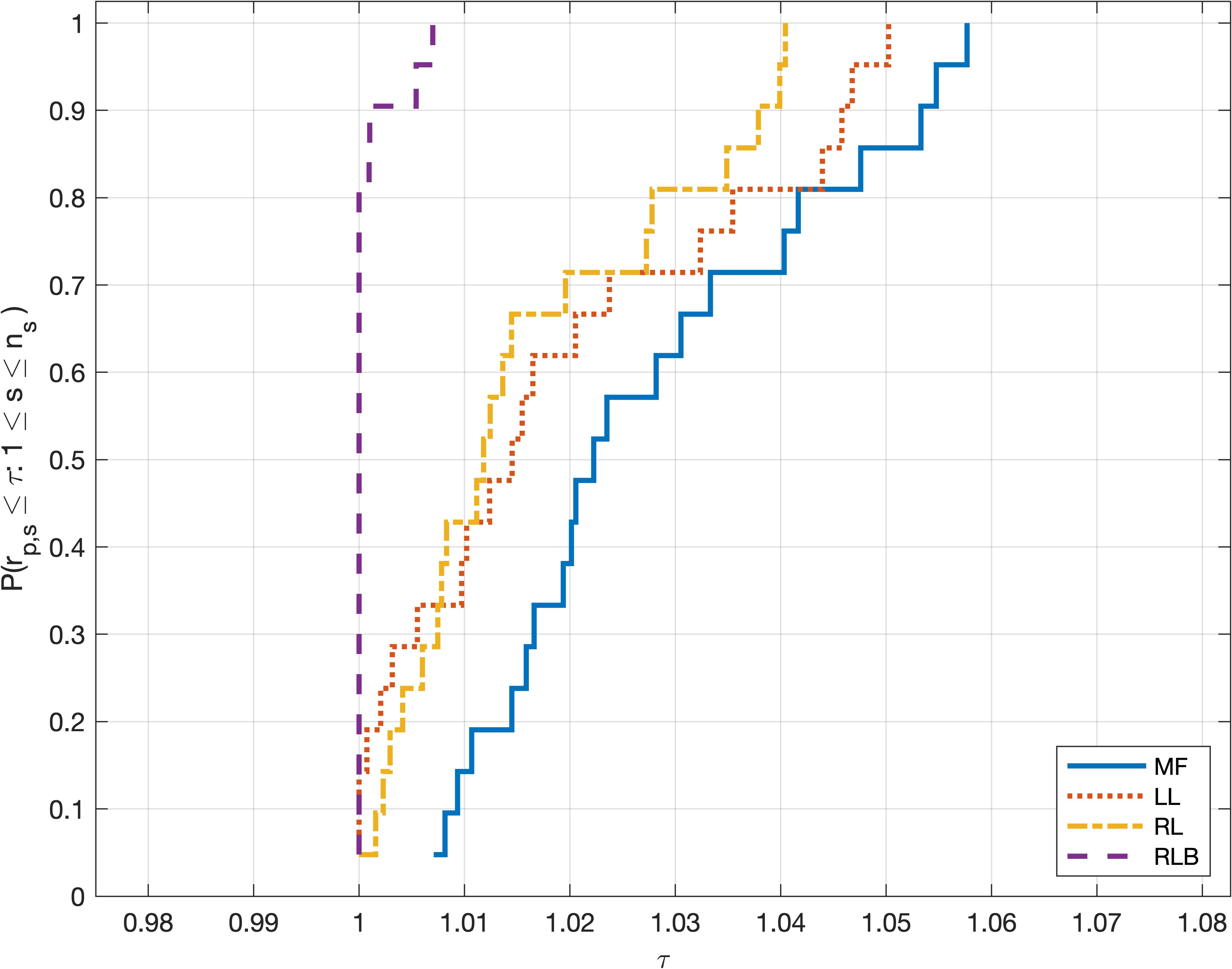}
  \label{fig:serial}
\end{center}
\end{figure}

We ran~MF, LL, RL, and~RLB on the large matrices,
with the serial BLAS linked in.
The performance profile for the factorization times
is shown in Figure~\ref{fig:serial}.
Unequivocally,~RLB is the fastest of the four methods;
it is the fastest method for eighty percent of the matrices,
and it is within one percent of the fastest time
for the remaining twenty percent of the matrices.

Clearly,~RL is second fastest overall, and~MF is fourth fastest overall.
This supports our earlier contention that~RL is modestly faster than~MF.
It is the case, however, that for each of the matrices,
the slowest time is less than a six percent increase over the
fastest time.
Therefore, we are dealing with fairly modest differences in the runtimes whenever the 
serial BLAS are linked in.

Next, we ran~MF, LL, RL, and~RLB on the large matrices,
with the multithreaded BLAS linked in.
All of the runs used~48 cores,
which is the maximum available.
(We tested how the methods scaled, and we
obtained our best timings using~48 cores.)
Because the timings for these factorizations were
significantly more unstable than those obtained using the serial BLAS,
we repeated the factorizations seven times
and then used the median timing.

\begin{figure}[htb]
\begin{center}
  \caption{Performance profile for the factorization times for the~21 large matrices
           whenever the multithreaded BLAS are linked in.}
  \includegraphics[width=.8\textwidth]{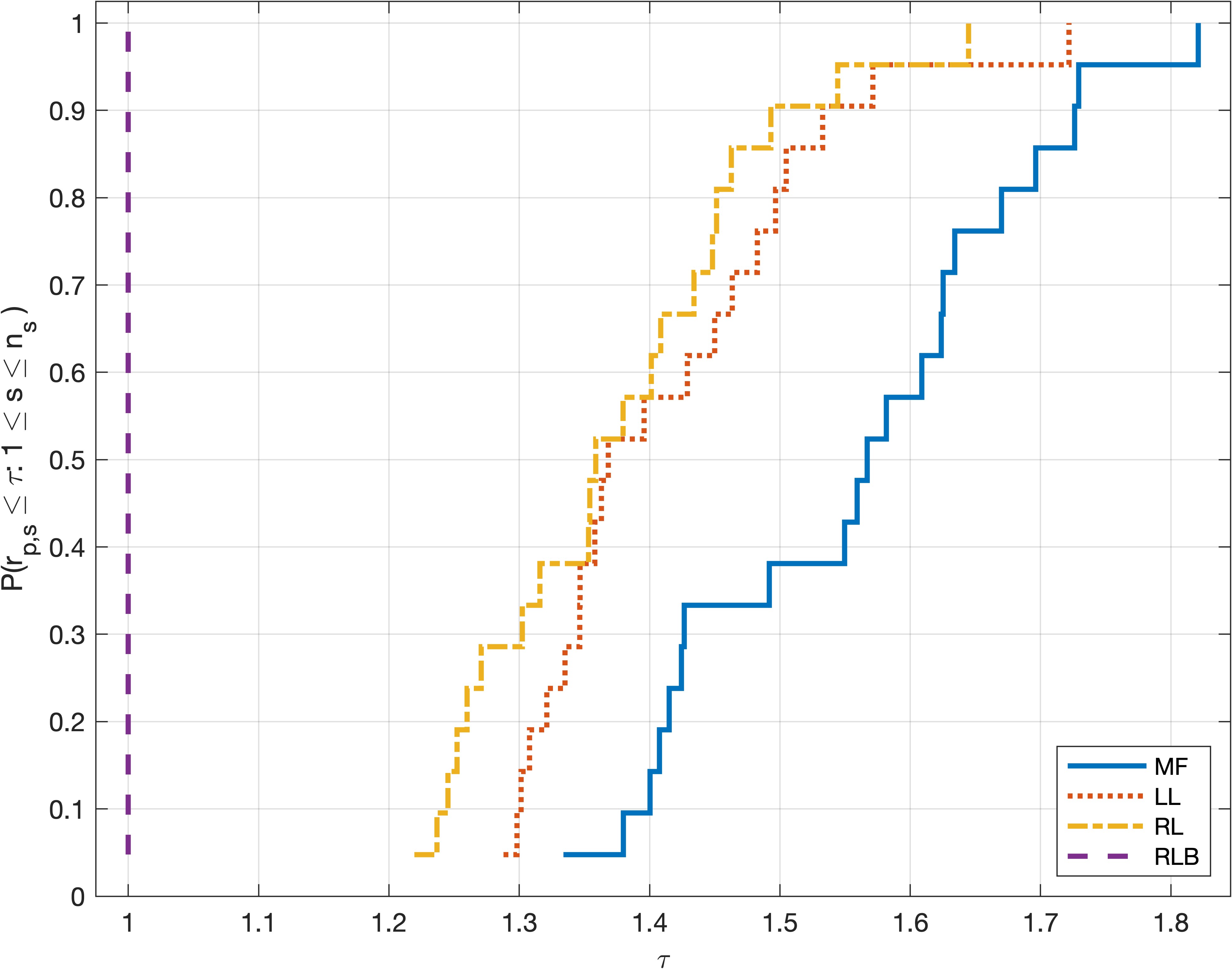}
  \label{fig:parallel}
\end{center}
\end{figure}

The performance profile for the factorization times
is shown in Figure~\ref{fig:parallel}.
Again, RLB is unequivocally the fastest method;
indeed, it is the fastest method for every matrix.
Here, we see that~RLB provides a very significant improvement
over the other methods.
In every case, using any one of the other methods
incurs an increase of between roughly twenty to eighty
percent in the runtime.
We conjecture that the performance of~LL, RL, and~MF suffers seriously
due to the fact that the assembly operations are performed serially.
In contrast, every floating-point operation in~RLB is performed
by Intel's MKL multithreaded library.

In this performance profile, there are also more substantial
differences in performance when comparing~LL, RL, and~MF.
Methods~LL and~RL track each other quite closely,
but both of these methods are substantially better than~MF.
We conjecture that the costs of the data movement
associated with~MF's stack (see line~18 in Algorithm~\ref{alg:MF})
are hurting the performance of~MF relative to~LL and~RL.
In any case,~RL is a greater improvement in speed over~MF
when multithreaded BLAS are used than it is when serial
BLAS are used.

\begin{figure}[htb]
\begin{center}
  \caption{Performance profile for the floating-point storage requirements
           for the~21 large matrices.}
  \includegraphics[width=.8\textwidth]{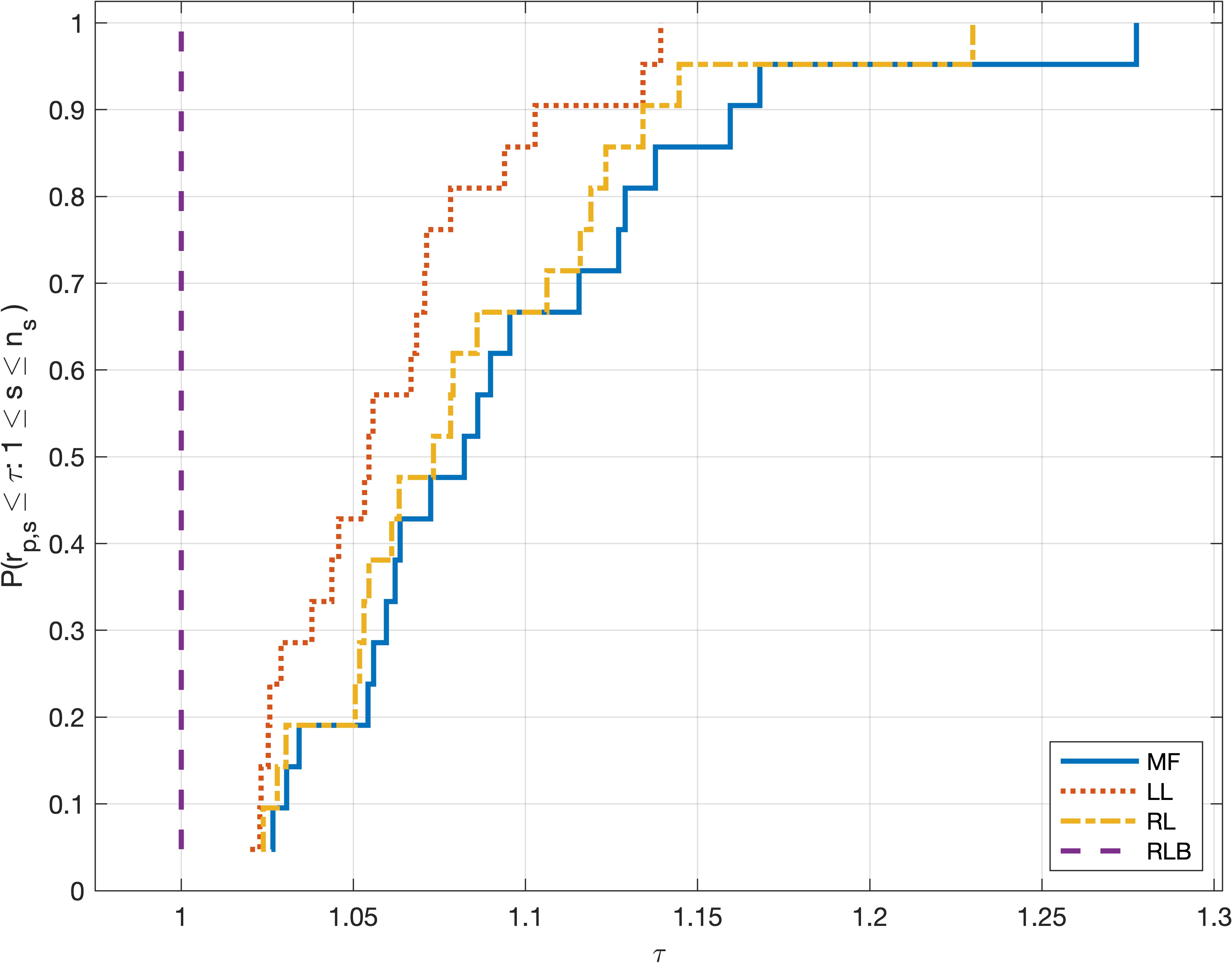}
  \label{fig:storage}
\end{center}
\end{figure}

All four factorization methods use exactly the same floating-point storage
for the factor matrix.
But~LL, RL, and~MF all require varying amounts of floating-point 
working storage for the update matrices;
RLB, of course, requires no additional floating-point working storage.
The total floating-point storage requirement is the sum of the two.
Clearly, RLB requires the least floating-point storage for every matrix.
The performance profile for the floating-point storage requirements
is shown in Figure~\ref{fig:storage}.
In most cases, the floating-point working storage increases the total
storage by less than ten percent.
Note that~LL is clearly the second best method;
note also that~RL is just marginally better than~MF.

We feel that the following sheds some light on the work presented
in this paper.
When~MF and~LL replaced Algorithm~\ref{alg:gsfct} years ago,
there was an increase in factorization efficiency
at some cost in floating-point storage.
Such trade-offs are very common in scientific computing in general.
In this study, we are {\em both} reducing floating-point storage
and increasing factorization efficiency by introducing~RLB.

At this point, we use speedups to evaluate the performance of~RLB
with the multithreaded BLAS lined in.
Recall that we made those runs using~48 cores.
Since~RLB with the serial~BLAS linked in proved to
be faster than its competitors,
we will compute the speedups relative to RLB.
In Table~\ref{tab:speedups}, the~21 large matrices
are listed in ascending order by factorization operations count.
\begin{table}[htb]
\begin{center}
\begin{tabular}{|l||r|r|r|} \hline
\multicolumn{1}{|c||}{} &
\multicolumn{2}{c|}{RLB time} &
\multicolumn{1}{c|}{} \\ \cline{2-3}
\multicolumn{1}{|c||}{} &
\multicolumn{1}{c|}{} &
\multicolumn{1}{c|}{Multi-} &
\multicolumn{1}{c|}{} \\
\multicolumn{1}{|c||}{Matrices} &
\multicolumn{1}{c|}{Serial} &
\multicolumn{1}{c|}{threaded} &
\multicolumn{1}{c|}{Speedup} \\ \hline
CurlCurl\_2        &    18.374  &    3.475  &    5.3   \\
dielFilterV2real   &    21.588  &    4.748  &    4.5   \\
dielFilterV3real   &    22.409  &    4.634  &    4.8   \\
PFlow\_742         &    25.717  &    3.333  &    7.7   \\
CurlCurl\_3        &    64.871  &    7.201  &    9.0   \\
StocF-1465         &    71.800  &    9.186  &    7.8   \\
bone010            &    74.247  &    7.031  &   10.6   \\
Flan\_1565         &    76.624  &    8.929  &    8.6   \\
audikw\_1          &   110.702  &    9.161  &   12.1   \\
Fault\_639         &   136.422  &    8.666  &   15.7   \\
Hook\_1498         &   160.296  &   13.064  &   12.3   \\
Emilia\_923        &   236.750  &   14.211  &   16.7   \\
CurlCurl\_4        &   249.682  &   19.578  &   12.8   \\
nlpkkt80           &   284.058  &   17.432  &   16.3   \\
Geo\_1438          &   331.670  &   19.857  &   16.7   \\
Serena             &   553.450  &   29.188  &   19.0   \\
Long\_Coup\_dt0    &   971.260  &   45.587  &   21.3   \\
Cube\_Coup\_dt0    &  1751.357  &   75.980  &   23.1   \\
Bump\_2911         &  3522.585  &  143.164  &   24.6   \\
nlpkkt120          &  4430.719  &  191.528  &   23.1   \\
Queen\_4147        &  4958.583  &  194.905  &   25.4   \\ \hline
\end{tabular}
\caption{RLB factorization times for the~21 large matrices using serial BLAS and then multithreaded BLAS,
         along with the associated speedups.
}
\label{tab:speedups}
\end{center}
\end{table}
We record RLB factorization times, using both serial and multithreaded BLAS,
and then the associated speedups.

It is no surprise that the speedups increase dramatically as the factors
become denser,
from the top of the table to the bottom.
As the factors become denser, the average block sizes increase
in both length and width,
improving the performance of the multihreaded DSYRK and DGEMM
operations.
Note that for~14 of the bottom~15 matrices in the table,
we have greater than an order of magnitude speedup.
The speedups exceed~20 for the bottom five matrices;
the efficiency is near 50 percent for the bottom four matrices.
Finally, for the  bottom five matrices,
the factorization is computed using the multithreaded BLAS
at rates exceeding one teraflop.
For Long\_Coup\_dt0, it is $1.13$ Tflops;
for Cube\_Coup\_dt0, it is $1.24$ Tflops;
for Bump\_2911, it is $1.32$ Tflops;
for nlpkkt120, it is $1.23$ Tflops; and
for Queen\_4147, it is $1.35$ Tflops.

%
%

\section{Conclusion}
\label{sec:conclusion}

In this paper, we have introduced two new variants of serial
sparse Cholesky factorization, RL and~RLB.
As we have seen, RL can be viewed as a variant of the multifrontal method~(MF).
In our testing and in our descriptions of the algorithms,
we confirmed that~RL is simpler than~MF;
we also confirmed that~RL is slightly faster than~MF whenever
Intel's MKL serial BLAS are linked in, and
we confirmed that~RL is significantly faster than~MF whenever
Intel's MKL multithreaded BLAS are linked in. 
We confirmed, moreover, that~MF requires marginally more
floating-point working storage than~RL.
The multifrontal method, however, has one very significant advantage over all of the
other factorization methods included in this study;
it is, by far, the best method for an implementation of
out-of-core factorization.

We believe that our findings regarding~RLB are of more interest
than anything else in the paper.
First, the use of~RLB must be preceeded by a careful
reordering within supernodes in order to increase
the size of the blocks
exploited by the method.
We use the method based on {\em partition refinement}~(PR)
to reorder within supernodes~\cite{JNP18,KNP24}.

Preceeded by the~PR reordering,
RLB outperforms, in every regard, the other factorization methods:
\begin{enumerate}
  \item
    It is the fastest whenever the serial BLAS are linked in.
  \item
    It is, by far, the fastest whenever the multithreaded BLAS are linked in.
  \item
    It is the fastest on the small matrices.
  \item
    It is the fastest on the large matrices.
  \item
    It uses the least floating-point storage.
\end{enumerate}
Finally, RLB, with the multithreaded BLAS linked in,
embodies the approach taken by LAPACK to run in parallel on machines with multiple cores.
We have shown that,
despite the sparsity in the factor matrices,
we can obtain good speedups using the same strategy that
LAPACK uses.
The key, here, is that RLB performs all of its floating-point operations within
the BLAS, 
while the other factorization methods perform assembly operations,
in addition to the operations that they perform within the BLAS.

\section*{ACKNOWLEDGMENTS}

This work was supported in part by the U.S. Department of Energy, Office of Science, Office of Advanced Scientific Computing Research and Office of Basic Energy Sciences, Scientific Discovery through Advanced Computing (SciDAC) Program through the FASTMath Institute and BES Partnership under Contract No. DE-AC02-05CH11231 at Lawrence Berkeley National Laboratory.

\bibliographystyle{ACM-Reference-Format}
\bibliography{mlf_paper}

\end{document}